\documentclass[11pt,twoside]{article}

\usepackage{asp2006}
\usepackage{epsf}
\usepackage{psfig}
\usepackage{lscape}
\usepackage{epsfig}

\begin{document}
\title{The Impact of Stellar Populations on the Dynamics of Merger Remnants}   
\author{B. Rothberg, J. Fischer}   
\affil{Naval Research Laboratory, Remote Sensing Division, Code 7211,
4555 Overlook Ave SW, Washington D.C. 20375}    

\begin{abstract} 
Many studies and simulations suggest gas-rich mergers do not contribute 
significantly to the overall star-formation rate and total mass function of galaxies.  The velocity 
dispersions ($\sigma$) of Luminous \& Ultraluminous Infrared Galaxies measured using the 1.62 or 2.29$\micron$ CO 
bandheads imply they will form m $<$ m$^{*}$ ellipticals.  Yet, $\sigma$'s obtained with the Calcium II 
triplet (CaT) at 0.85$\micron$ suggest all types of mergers will form m $>$ m$^{*}$ ellipticals.   Presented here are 
recent results, based on high-resolution imaging and multi-wavelength spectroscopy, which 
demonstrate the dominance of a nuclear disk of Red Supergiants (RSG) or Asymptotic Giant Branch (AGB) stars in the near-infrared 
bands, where dust obscuration does not sufficiently block their signatures. The presence of these stars severely biases 
the dynamical mass.  At {\it I}-band, where dust
can sufficiently block RSG or AGB stars, LIRGs populate the Fundamental Plane over a large
dynamic range and are virtually indistinguishable from elliptical galaxies.
\end{abstract}

\section{Introduction}  
\indent The ``Toomre Hypothesis'' \citep{1972ApJ...178..623T, 1977egsp.conf..401T},
proposes that the merger of two-gas rich spiral galaxies will form an elliptical galaxy, often with
a final stellar mass larger than the sum of the progenitors.  
In the local universe, Luminous and Ultraluminous Infrared Galaxies are ideal candidates for 
forming massive elliptical galaxies \citep{1992ApJ...390L..53K}.  These are objects with 
{\it L}$_{\rm IR}$ $>$ 10$^{11}$ L$_{\odot}$ between 8-1000 $\micron$ 
\citep{1996ARAA..34..749S}, contain vast quantities of molecular gas, and show strong evidence of recent 
or ongoing merging activity.  Radio recombination line observations of the nearest ULIRG Arp 220 
imply a formation rate of 10$^{3}$ M$_{\odot}$ yr$^{-1}$ \citep{2000ApJ...537..613A}, while CO 
interferometric data indicate that 0.15-0.46 of the dynamical mass of this system is gaseous 
\citep{1998ApJ...507..615D,2009ApJ...692.1432G}. The star-formation rates and vast 
quantities of gas in LIRGs/ULIRGs could add a significant stellar component to the total mass of the merger.\\
\indent However, a number of studies, all using infrared CO bandheads to measure central velocity dispersions
($\sigma$$_{\circ}$), have shown that LIRGs have masses consistent with low-moderate luminosity elliptical galaxies 
({\it L} $\sim$ 0.03-0.15 {\it L}$^{*}$) \citep{1998ApJ...497..163S} and ULIRGs have masses
consistent with {\it L} $\leq$ {\it L}$^{*}$ (e.g. \cite{2001ApJ...563..527G}). These results
have raised significant doubts as to whether gas-rich mergers contribute significantly to the formation of elliptical galaxies.\\
\indent Yet, $\sigma$$_{\circ}$ measured from the Calcium II Triplet absorption lines (CaT),
suggest gas-rich mergers, including LIRGs, have masses which span nearly the entire mass range of
elliptical galaxies \citep{1986ApJ...310..605L,2006AJ....131..185R} (hereafter RJ06).  This difference in $\sigma$,
or $\sigma$-mismatch, is counter-intuitive.  Namely, that LIRGs/ULIRGs, which are
undergoing intense star-formation and possess large quantities of dust, should show smaller $\sigma$$_{\circ}$
at {\it longer} wavelengths.  The use of infrared stellar lines to measure $\sigma$$_{\circ}$ was initially
motivated by the need to pierce the veil of extinction in starburst galaxies and measure their ``true'' dynamical masses.
However, the results presented here show that IR-luminous mergers are Janus-like, that is, they reveal 
two different dynamical faces depending on the wavelength observed.  The obscuring characteristics of dust in
the optical in IR-luminous galaxies behaves in a manner beneficial for determining the true mass of merger remnants.

\section{Sample Selection \& Observations}
\indent The dynamical properties of a sample of 14 advanced (single-nuclei) 
merger remnants are compared with a sample of 23 elliptical galaxies.   The merger remnants 
are a subsample of the 51 merger remnants discussed in detail in \cite{2004AJ....128.2098R} (hereafter RJ04). 
The photometric data for the merger remnants include {\it F814W} ($\sim$ {\it I}-band) imaging from the 
Wide-Field Planetary Camera 2 ({\it WFPC2}) or the Advanced Camera for Surveys Wide-field Camera ({\it ACS/WFC})
on {\it HST} and {\it K}-band imaging from Quick Infrared Camera (QUIRC) on the University of 
Hawaii 2.2m telescope.  The kinematic data for the merger remnants include CaT observations from ESI on Keck-2,
and CO observations from either NIRSPEC on Keck-2 or GNIRS on Gemini South (Program GS-2007A-Q-17, P.I. Rothberg).
Additional kinematic and photometric data were obtained from the literature for 
several merger remnants and the comparison sample of ellipticals.

\section{The Fundamental Plane and Stellar Populations}
{\begin{figure}[!ht]
\epsfig{file=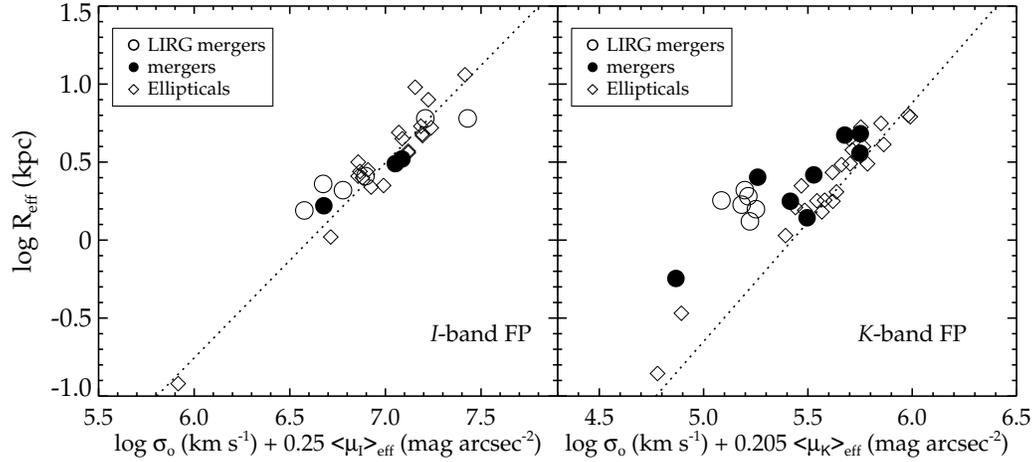,scale=0.7}
\caption{Shown are the {\it I}-band ({\it left}) and {\it K}-band ({\it right}) Fundamental Planes
from \cite{1997AJ....113..101S} and \cite{1998AJ....116.1591P} respectively. Overplotted in both 
panels are LIRG merger remnants (open circles), non-LIRG merger remnants (filled circles) and 
elliptical galaxies (open diamonds).   All 6 LIRGs, 20/23 ellipticals, and 3/8 non-LIRG merger 
remnants are overplotted on the {\it I}-band FP. }
\end{figure}
}
\indent A similar $\sigma$-mismatch was reported for a sample of 25 nearby early-type (predominantly S0) galaxies
by \cite{2003AJ....125.2809S}.  They also found that $\sigma$$_{\circ, optical}$ $>$ $\sigma$$_{\circ, CO}$, and suggested
that the kinematics of the cold stellar component in S0 galaxies was obscured by dust, and detectable only in the IR
while the kinematics of the hot spheroid dominated optical wavelengths.   However,
it remained unclear whether bonafide ellipticals produced the same discrepancy.
As noted in \cite{2009arXiv0902.1725R}, no discernible difference between $\sigma$$_{\circ, optical}$ and $\sigma$$_{\circ, CO}$
was found for the comparison sample of 23 elliptical galaxies.  On the other hand, LIRGs showed a large discrepancy,
as first noted in RJ06. The Fundamental Plane (FP) is a two-dimensional plane embedded in the three-dimensional parameter space 
of $\sigma$$_{\circ}$, the half-light (effective) radius ({\it R}$_{\rm eff}$), and the surface brightness within the effective radius 
($<$$\mu$$>$$_{\rm eff}$). {\it All} elliptical galaxies lie on the Fundamental Plane.  The LIRG/ULIRG studies noted earlier found
that while these mergers were overly luminous in the infrared, they would eventually evolve onto the FP, but the small $\sigma$$_{\circ, CO}$ 
meant they could not be the progenitors of ellipticals with {\it L} $>$ {\it L}$^{*}$.  RJ06, however, found that the 
observed range of $\sigma$$_{\circ, CaT}$ meant that gas-rich mergers could populate nearly all of the mass-range of ellipticals,
including {\it L} $>$ {\it L}$^{*}$.   Figure 1 is a two panel figure which shows the {\it I}-band and {\it K}-band FPs, 
with LIRG merger remnants (open circles), non-LIRG merger remnants (filled circles) 
and  ellipticals (open diamonds).    \\
\indent As expected, the ellipticals show little difference between the {\it I} and {\it K}-band FPs.  
The difference in the location of (primarily) the LIRGs in the {\it I}-band and {\it K}-band is striking.  
Figure 1 explains the apparent contradictory results between earlier LIRG/ULIRG studies, which used ``pure'' {\it H} or {\it K}-band
FPs and those from RJ06, which used a ``hybrid'' FP (CaT $\sigma$$_{\circ}$ and {\it K}-band photometry).
In the {\it I}-band, the dynamical properties of LIRGs are indistinguishable from ellipticals.  Figure 2 shows a comparison between
the {\it M$_{\rm dyn}$/L} and {\it M$_{\rm dyn}$} in the {\it I}-band ({\it left}) and {\it K}-band ({\it right}) for the merger remnants
and elliptical galaxies (same symbols as Figure 1).  Overplotted is the evolution of {\it M/L} over time for a burst population
from \cite{2005MNRAS.362..799M} (hereafter M05).  Figure 2 shows that in the {\it I}-band, the measured dynamical masses and stellar ages 
of the LIRGs are nearly the same as elliptical galaxies.  However, the {\it K}-band measurements imply LIRGs have smaller 
{\it M}$_{\rm dyn}$ and young ages. \\ 
\indent The {\it K}-band is dominated by the presence of young stars.  Numerical simulations have long predicted that gaseous dissipation
in the merging event funnels the gas into the barycenter of the merger (e.g. \cite{1991ApJ...370L..65B,2002MNRAS.333..481B}).
This forms a rotating gaseous disk in the central 1-2 kpc of the merger, which then undergoes a strong starburst, forming a rotating
disk of young stars.  One observational signature of this starburst is the presence of ``excess light'' in the surface brightness
profiles of mergers (e.g. \cite{1994ApJ...437L..47M}), first detected in the {\it K}-band by RJ04.  The $\sigma$-mismatch 
detected in IR-luminous galaxies is another observational signature of these rotating central starbursts.  Dust
associated with these nuclear starbursts blocks most of their light at $\lambda$ $<$ 1$\micron$, while allowing the 
random motions of the nearly virialized older stars to dominate the $\sigma$$_{\circ}$
measurement at {\it I}-band.  This functions in a similar manner to an occulting mask in a coronograph.   At {\it H} and {\it K}-band, the 
RSG and AGB stars can account for 60-90$\%$ of the light (M05), therefore 
the disk kinematics overwhelms the $\sigma$$_{\circ, CO}$ measurement.

{\begin{figure}[!ht]
\epsfig{file=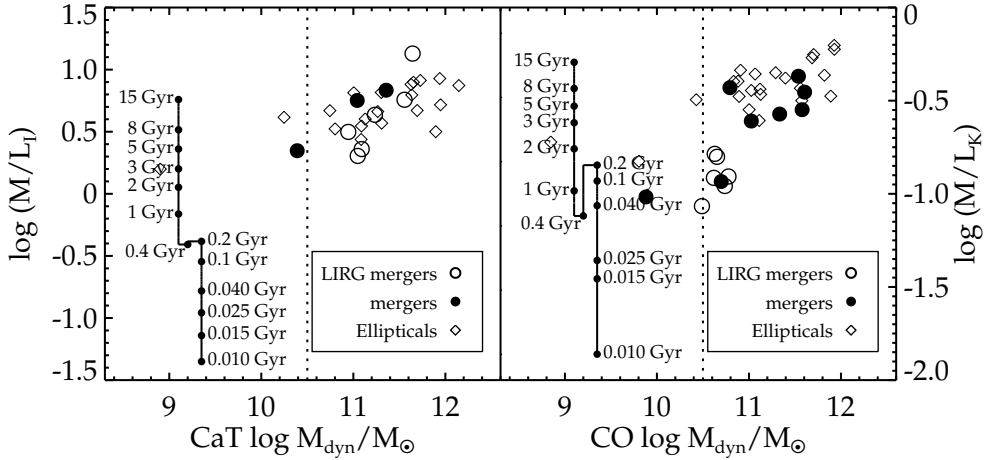,scale=0.9}
\caption{Two panel figure showing pure {\it I}-band ({\it left}) and pure {\it K}-band ({\it right}) {\it M}/{\it L}
vs. {\it M}$_{\rm dyn}$.  The overplotted vector (solid line) in each panel is the evolution of {\it M}/{\it L} for a single-burst
stellar population with solar metallicity and a Salpeter IMF as computed from \cite{2005MNRAS.362..799M}.  The dotted vertical
line in each panel indicates {\it m}$^{*}$.}
\end{figure}
}


\end{document}